\documentclass[twocolumn,english,aps,prx,reprint]{revtex4-2}
\usepackage[T1]{fontenc}
\usepackage[latin9]{inputenc}
\usepackage{babel}
\usepackage{amsmath}
\usepackage{amssymb}
\usepackage{graphicx}
\usepackage[bookmarks=false,
 breaklinks=false,pdfborder={0 0 1},backref=false,colorlinks=false]
 {hyperref}

\makeatletter
\usepackage{babel}
\usepackage{physics}

\makeatother

\begin{document}
\title{Which Rope Breaks? A Study of Tension Distribution in Multi-Rope Systems}
\author{Amir Eskandari-asl$^{1,*}$ and Roberto De Luca$^{1}$}
\address{$^{1}$Dipartimento di Fisica ``E.R. Caianiello'', Universit� degli
Studi di Salerno, I-84084 Fisciano (SA), Italy}
\begin{abstract}
We investigate the tension distribution in systems of mass-less ropes
under different loading conditions. For a two-rope system, we demonstrate
how the breaking scenario depends on the applied force dynamics: rapid
pulling causes the lower rope to break, while gradual pulling leads
to upper rope failure. Extending to a three-rope Y-shaped configuration,
we identify a critical angle $\theta_{C}=60^{\circ}$ that determines
which rope breaks first. When the angle between the upper ropes exceeds
this critical value, the upper ropes fail before the lower one. We
further analyze how an attached mass at the junction point modifies
this critical angle and establish maximum mass limits for valid solutions.
Our results provide practical insights for introductory physics students
understanding static forces and system stabilities.
\end{abstract}
\maketitle

\section{Introduction}

The study of tension and stability in ropes, beyond its practical
importance which motivates extensive research works \citep{guo2020experimental,chen2021wire,mao2023data,xu2023nn,zhang2023vision},
offers an engaging entry point for introducing students to fundamental
principles of classical mechanics, particularly Newton's laws and
force balance. Simple rope configurations can reveal complex behaviors
that challenge intuition and prompt deeper investigation into the
role of dynamics and geometry in force distribution.

In this paper, we analyze two common configurations: a vertical two-rope
system and a Y-shaped three-rope system. These setups, while conceptually
accessible, yield rich physical insights when carefully examined,
and help the students overcome some difficulties in understanding
basic concepts \citep{flores2010students}. For the two-rope system,
we show how the sequence of rope failure depends on how quickly it
is applied, highlighting the role of acceleration and inertial effects.
In the three-rope system, we identify a critical angle beyond which
the upper ropes fail under quasi-static loading.

Beyond theoretical interest, these configurations lend themselves
well to classroom demonstrations and laboratory experiments. Using
readily available materials, such as identical ropes, spring scales,
pulleys, and variable masses, students can recreate the described
systems and test the predicted outcomes. In the three-rope system,
by varying the angle between ropes, students can observe directly
which rope breaks first and can correlate their findings with theoretical
predictions. The addition of a mass at the junction point introduces
further variables, offering opportunities for experimental design
and data analysis.

These experiments not only reinforce elementary physics concepts but
also encourage scientific reasoning, model testing, and collaborative
problem-solving. We hope this work will trigger the interest of physics
teachers and their students at high-school or undergraduate university
level.

\section{Two Ropes\protect\label{sec:2ropes}}

\begin{figure}[t]
\centering{}\includegraphics[width=7cm]{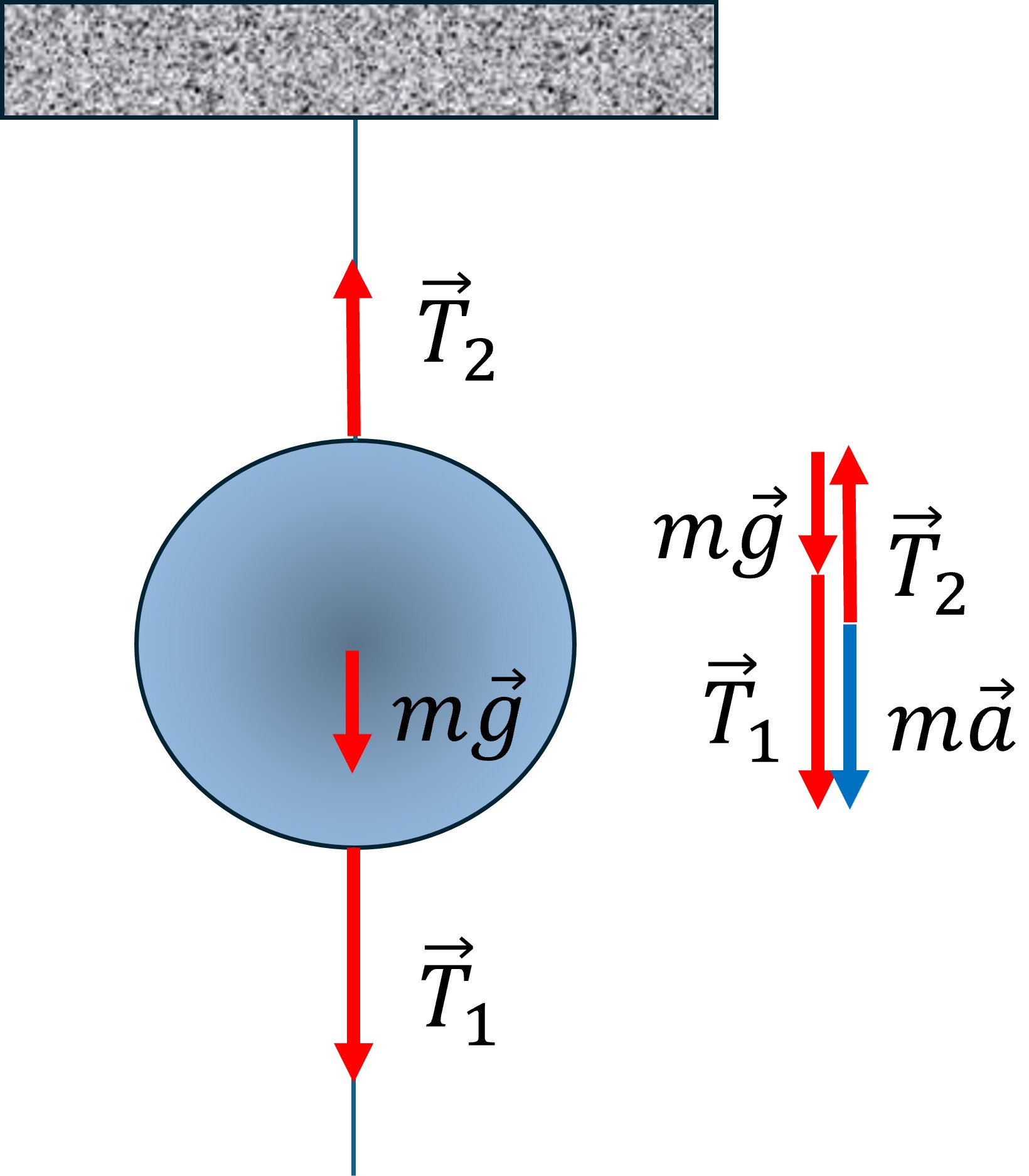}\caption{Two-rope configuration: A mass $m$ connects an upper rope (tension
$T_{2}$) and a lower rope (tension $T_{1}$). The forces are also
presented in a polygon diagram. The breaking behavior depends on how
force is applied to the lower rope. For a rapid pull down, the ropes
cannot be considered rigid, and the mass accelerates for a short period
of time. However, for a static situation with the force $\protect\overrightarrow{T}_{1}$
gradually increasing, the acceleration, $\protect\overrightarrow{a}$,
vanishes and we assume the ropes to be ideally rigid.\protect\label{fig1:2ropes}}
\end{figure}

Consider a rope suspended from a fixed point, with a mass $m$ attached
to its free end. When we attach a second rope to the mass and leave
its other end free, we obtain the configuration shown in Fig. \ref{fig1:2ropes}.
Both ropes are identical, mass-less, defect-free, and completely rigid
with no elasticity (we will temporarily relax the latter condition
and again consider it in the rest of the manuscript). We examine what
happens when we pull down the lower rope: which rope breaks first
depends on how the force is applied.

As Halliday and Resnick discuss \citep{halliday2014fundamentals},
two scenarios emerge. In the first scenario, where the lower rope
is pulled rapidly, it breaks while the upper rope remains intact.
In the second scenario, with gradual pulling, the upper rope breaks
first.

The physics behind these scenarios can be understood by noting that
in a real setup we don't deal with an ideal completely rigid rope.
Indeed, the ropes have a finite elasticity and hence upon a rapid
pull down, we can assume the mass to be accelerated for a very short
time period. At a time instant during that period, one can use the
Newton's second law for the attached mass to write (see Fig. \ref{fig1:2ropes}):

\begin{equation}
T_{1}-T_{2}=m\left(a-g\right).
\end{equation}

For the rapid pulling we have $a\gg g$ and hence, $T_{1}\gg T_{2}$.
Consider the breaking force threshold, $T_{B}$, such that if the
tension force exceeds this value, the rope would break. During the
rapid pull down, when $T_{1}$ reaches the breaking tension $T_{B}$,
$T_{2}$ remains below $T_{B}$, causing the lower rope to break.

On the other hand, for gradual pulling there is no acceleration and
we have $a\approx0$. Hereafter, we work in this regime (gradual pulling),
and retain our idealization of fully rigid non-elastic ropes. Applying
the Newton's second law for this case , we find:

\begin{equation}
T_{2}=T_{1}+mg>T_{1}.
\end{equation}

Here, gradually pulling down the lower rope, $T_{2}$ reaches $T_{B}$
first, because it is larger than $T_{1}$, and therefore the upper
rope is the one which breaks. As mentioned above, these results assume
identical breaking tensions $T_{B}$ for all ropes.

\begin{figure}[t]
\centering{}\includegraphics[width=7cm]{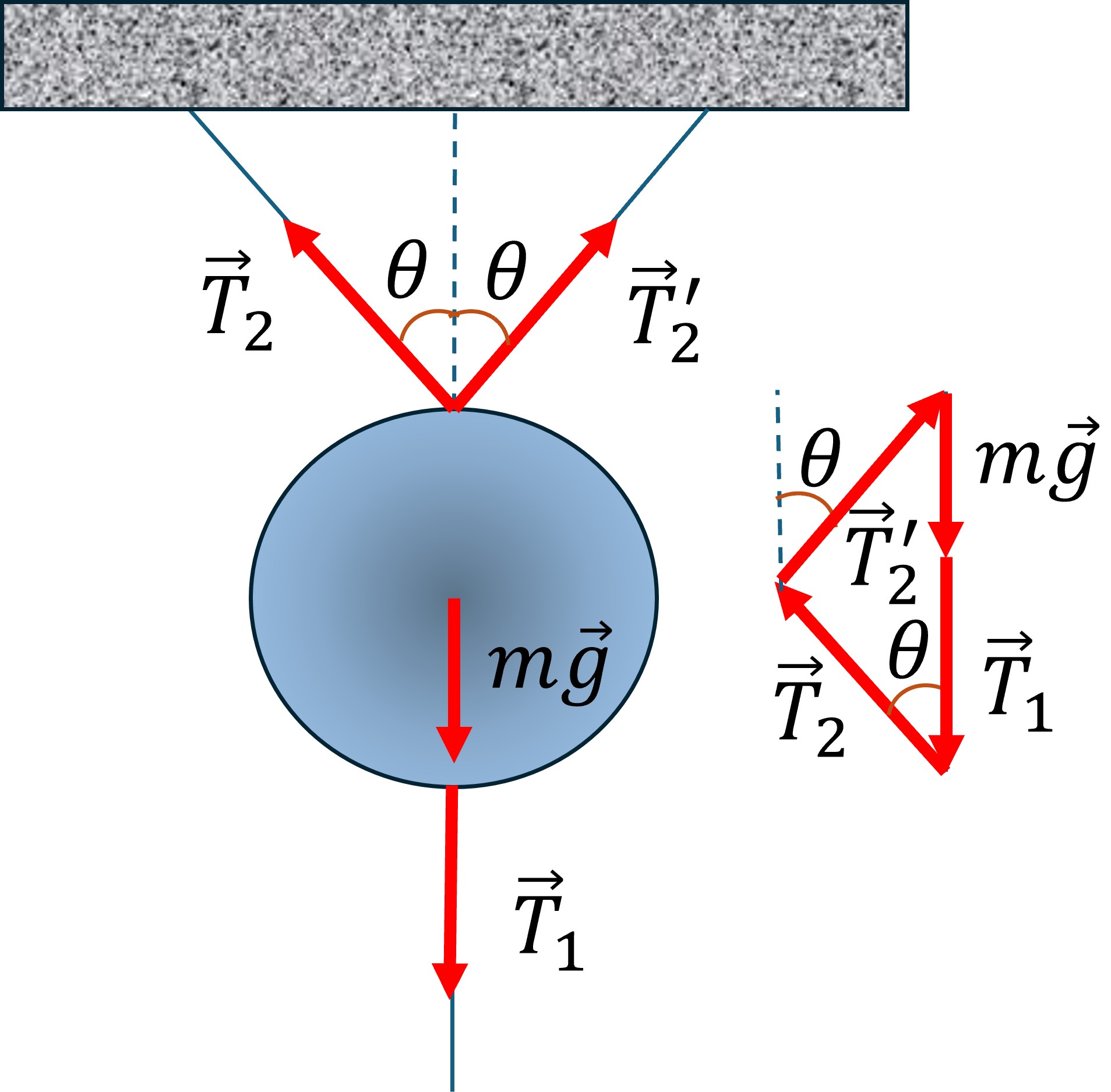}\caption{Three-rope Y-configuration with the forces shown in polygon too. The
angle $2\theta$ between upper ropes determines which rope breaks
first under vertical loading. The value of the tension force in the
upper ropes is the same: $T_{2}=T_{2}^{\prime}$.\protect\label{fig2:3ropes}}
\end{figure}

\section{Three Ropes and The Critical Angle\protect\label{sec:3ropes_direct}}

The problem becomes more intriguing with three ropes arranged in a
Y-shape, as shown in Fig. \ref{fig2:3ropes}. The upper two ropes
form an angle $2\theta$ and connect to a third vertical rope. In
what follows we shall only consider the gradual pulling scenario.

Based on intuition and comparing to the previous case, one expects
that when the attached mass is small enough ( to be more precise,
when the gravity force is less than the breaking tension threshold,
$T_{B}$) and the upper ropes are vertical (i.e., for $\theta=0$),
always the lower rope breaks. However, the question is, what happens
if $\theta>0$ ? This issue is addressed in the following sections.

\subsection{No mass attached}

First, consider a mass-less connection point ($m=0$). Force balance
gives: 
\begin{equation}
T_{1}-2T_{2}\cos\theta=0\quad\text{or}\quad\frac{T_{1}}{T_{2}}=2\cos\theta.\label{eq:2rope-mass0}
\end{equation}

This reveals a critical angle, for which $T_{1}=T_{2}$: 
\begin{equation}
\theta_{C}=\cos^{-1}(\frac{1}{2})=60^{\circ}.
\end{equation}

For $\theta<\theta_{C}$, $\cos\theta>1/2$ leads to $T_{1}>T_{2}$,
so the lower rope breaks first. The extreme case of the vertical upper
ropes, $\theta=0^{\circ}$ gives $T_{1}=2T_{2}$, clearly breaking
the lower rope, as one would expect.

For $\theta>\theta_{C}$, $T_{1}<T_{2}$ holds, so the upper ropes
break first. This occurs because at larger angles, $\cos\theta<1/2$
and therefore $T_{2}$ is larger than the sum of the vertical components,
$2T_{2}\cos\theta$. These conclusions are independent of $T_{B}$
value under our assumptions.

For $\theta=\theta_{C}$ the three ropes break at the same time in
an ideal situation. However, in reality the ropes are not perfectly
ideal and because of the defects in the ropes and/or errors in measuring
the angles, usually just one of the three ropes breaks.

This shows that in order to break a rope by a small vertical force,
one needs to tie up its ends, so that the angle created in the Y-shape
is close to $90^{\circ}$ and the rope can be broken by a smaller
force $T_{1}$. In a real setup though, the ropes have a little bit
of elasticity and therefore upon exerting the force $T_{1}$, the
angle $\theta$ gets smaller. In other words, the elasticity tends
to save the upper ropes and instead prefers the lower one to be broken.

One question that can be answered using the above discussion is, what
is the maximum force that can be exerted on this system? Or, in other
words, what exerted force would break this system? Here, we consider
the force to be exerted on the lower rope, and hence, is equal to
the its tension. The maximum tension above which the system would
break is dubbed $T_{1}^{(B)}$ in the following. If $\theta<\theta_{C}$,
the lower rope is the one which breaks, and hence, $T_{1}^{(B)}$
is just equal to the breaking tension, $T_{B}$. If $\theta>\theta_{C}$,
the upper ropes break, and hence the maximum exerted force is obtained
by letting $T_{2}$ reach its maximum value, which is $T_{B}$. Setting
$T_{2}\rightarrow T_{B}$ in Eq. \ref{eq:2rope-mass0}, one obtains
$T_{1}^{(B)}=2T_{B}\cos\theta$. This discussion is summarized in
the following equation:
\begin{equation}
T_{1}^{(B)}=\begin{cases}
T_{B}; & \theta<\theta_{C},\\
2T_{B}\cos\theta; & \theta\geq\theta_{C}.
\end{cases}\label{eq:T1B}
\end{equation}

\subsection{The effects of the attached mass}

In the previous section we considered the case where the mass of the
junction was negligible. Here, let's consider a case in which a weight
is attached to the junction. With mass $m$ at the junction ($m\neq0$)
and gradual pulling ($a=0$), the balance becomes: 
\begin{equation}
\frac{T_{1}+mg}{T_{2}}=2\cos\theta.\label{eq:theta_m}
\end{equation}
In this case, even if $\theta$ is smaller than the critical angle
of the zero-mass case, it doesn't necessarily mean the lower rope
will break, because the force $T_{2}$ should be larger than the zero-mass
case to compensate for $mg$. As such, the critical angle depends
on the mass and the breaking tension, $T_{B}$. The critical angle,
$\theta_{C,m}$, is obtained by setting $T_{1}=T_{2}=T_{B}$ in Eq.
\ref{eq:theta_m} as 
\begin{equation}
\theta_{C,m}=\cos^{-1}\frac{1+mg/T_{B}}{2},\label{eq:theta_Cm}
\end{equation}
which now depends on the value of $T_{B}$. Note that if $T_{B}\gg mg$,
we get back to the mass-less case. This indicates that the mass-less
contact point is a case in which the weight of the contact point is
much less than the breaking tension of the ropes.

Note that the weight $mg$ cannot be infinitely large. The maximum
value of the mass for each angle $\theta$ is the mass which would
just break the upper ropes without any force from the lower rope.
Setting $T_{1}=0$ and $T_{2}=T_{B}$ in Eq. \ref{eq:theta_m} we
get: 
\begin{equation}
\frac{m_{\text{max}}g}{T_{B}}=2\cos\theta.
\end{equation}
Therefore, noting $\cos\theta\leq1$, the mass we suspend should be
always small enough to satisfy 
\begin{equation}
\frac{m_{\text{max}}g}{T_{B}}<2.
\end{equation}

The above discussion clarifies that we have two different regimes,
considering the value of $\frac{mg}{T_{B}}$. The first regime is
when $\frac{mg}{T_{B}}\leq1$. In such a case, the argument of the
inverse-cosine function in Eq. \ref{eq:theta_Cm} is not larger than
unity, and hence the critical angle is well-defined. For $\theta\gtrless\theta_{C,m}$,
given that our angles are in the range $0^{\circ}$ to $90^{\circ}$,
we have $\cos\theta\lessgtr\cos\theta_{C,m}$. Substituting the values
of the cosine functions from Eqs. \ref{eq:theta_m} and \ref{eq:theta_Cm},
we get: 
\begin{equation}
T_{1}+mg\lesseqqgtr T_{2}+T_{2}\frac{mg}{T_{B}},\qquad\textrm{for }\theta\gtreqqless\theta_{C,m}.\label{ineq:theta_theta_C}
\end{equation}
Let's assume we pull down the system until $T_{2}\rightarrow T_{B}$.
For $\theta<\theta_{C,m}$, the inequality \ref{ineq:theta_theta_C}
implies that $T_{1}>T_{B}$, which is not possible, because the tension
in any rope cannot exceed $T_{B}$. Consequently, $T_{2}$ cannot
reach $T_{B}$ before $T_{1}$ does, and hence the lower rope breaks
and not the upper ones. On the other hand, if $\theta>\theta_{C,m}$,
the inequality \ref{ineq:theta_theta_C} implies that $T_{1}<T_{B}$,
which means that the upper rope breaks while the lower one remains
intact. This argument shows the same concept of the critical angle
as the case of the mass-less contact: if the angle $\theta$ is smaller
than $\theta_{C,m}$, the lower rope breaks first, while if it is
larger than $\theta_{C,m}$, the upper ones break first.

Similar to the previous section, we can find the exerted force which
would break the system. Clearly, for $\theta<\theta_{C,m}$, we just
have $T_{1}^{(B)}=T_{B}$, while for $\theta>\theta_{C,m}$ in which
case the upper rope breaks, we should set $T_{2}\rightarrow T_{B}$
in Eq. \ref{eq:theta_m}. The result is the following:
\begin{equation}
T_{1}^{(B)}=\begin{cases}
T_{B}; & \theta<\theta_{C,m},\\
2T_{B}\cos\theta-mg; & \theta\geq\theta_{C,m}.
\end{cases}\label{eq:T1Bm}
\end{equation}

The case $\frac{mg}{T_{B}}=1$ deserve a specific attention. In this
case, Eq. \ref{eq:theta_Cm} gives $\theta_{C,m}=0^{\circ}$. For
a non-zero $\theta$, being larger than the critical angle, the above
reasonings clarify that the upper ropes break upon pulling down the
system. For $\theta=\theta_{C,m}=0^{\circ}$, the inequality \ref{ineq:theta_theta_C}
tells us that if $T_{2}\rightarrow T_{B}$, also $T_{1}\rightarrow T_{B}$,
and therefore the three ropes break together, in ideal conditions.

The second regime is when $1<\frac{mg}{T_{B}}<2$. In this regime,
the argument of the inverse cosine function in Eq. \ref{eq:theta_Cm}
exceeds unity, and hence we don't have any critical angle. Using Eq.
\ref{eq:theta_m} and the fact that $\cos\theta\leq1$, we get $T_{1}+mg\leq2T_{2}$.
However, in this case, $T_{B}<mg$ and no tension can exceed $T_{B}$,
which means $T_{1}<mg$, and consequently, $2T_{1}\leq T_{1}+mg$.
Combining the two inequalities we get $T_{1}<T_{2}$, which shows
that by exerting the force, the tension in the upper ropes reach $T_{B}$
first, and independently of angle $\theta$, always the upper ropes
break. In short, in this large-mass regime, the upper ropes break.

\begin{figure*}[t]
\centering{}\includegraphics[width=18cm]{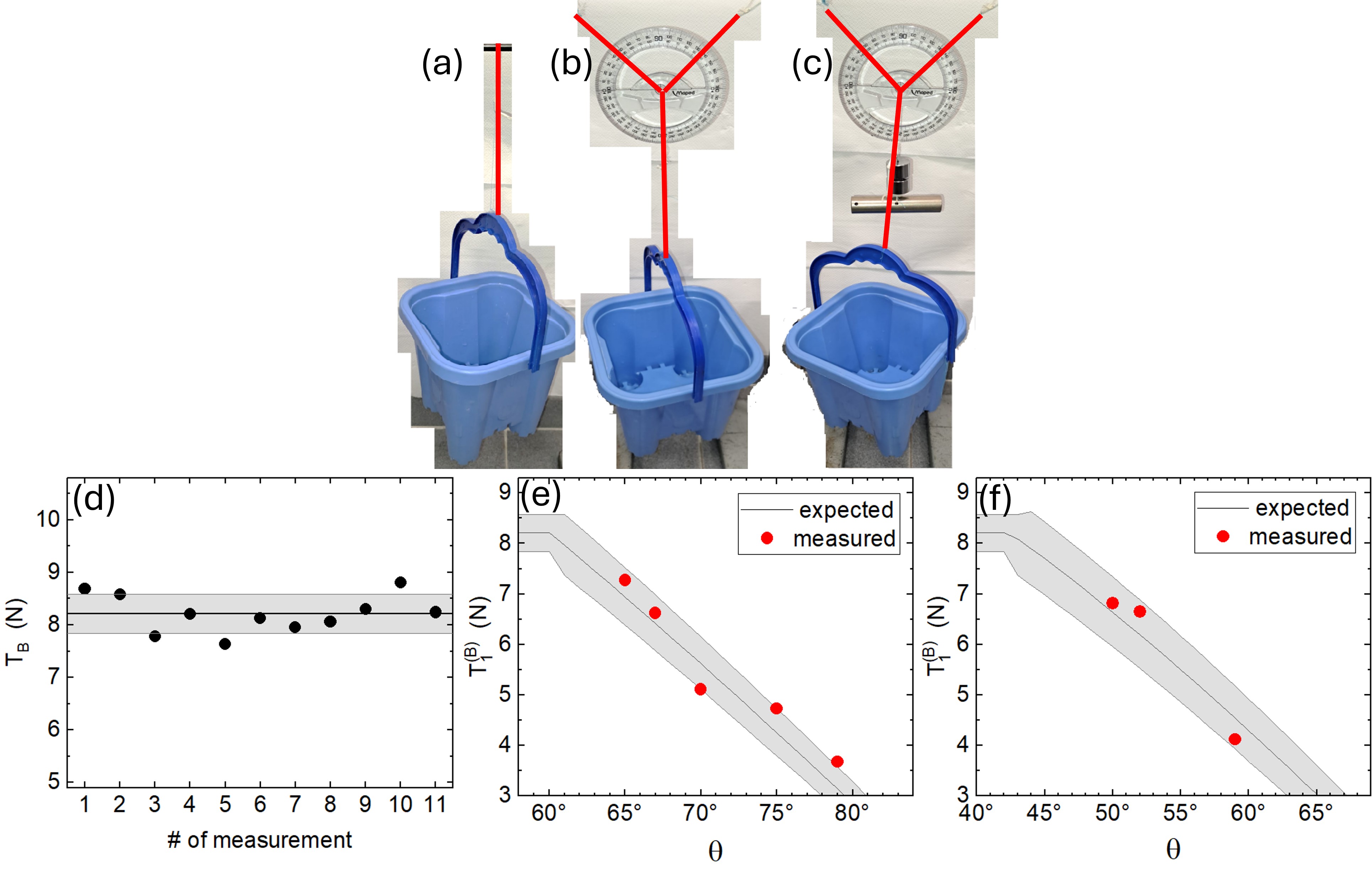}\caption{Experimental setup and results. (a) Measurement of breaking tension
$T_{B}$ using a single string. (b) Three-rope case (Y-configuration)
without mass at junction for measuring $T_{1}^{(B)}$. (c) Three-rope
case (Y-configuration) with mass $m=0.400$ kg at the junction. In
the panels (a),(b) and (c), the cotton strings are very difficult
to be seen as they are very thin, and are therefore marked by thick
red lines. Note that, due to the camera perspective, some of the vertical
strings may not appear perfectly vertical. (d) Measured $T_{B}$ values
from multiple trials, with gray band showing mean $\pm$ standard
deviation. (e) Maximum exerted force, $T_{1}^{(B)}$, versus angle
$\theta$ for massless three-rope case ($m=0$), compared with theoretical
prediction (gray band). (f) $T_{1}^{(B)}$ versus $\theta$ for the
three-rope case with mass at junction, showing reduced critical angle.\protect\label{fig3:experiment}}
\end{figure*}

\section{Experimental results}

To validate our theoretical predictions, we designed a simple experimental
setup using cotton strings instead of ropes, as they can be broken
with relatively small forces. All experiments were conducted under
quasi-static loading conditions, with forces applied gradually to
ensure negligible acceleration effects.

We first determined the breaking tension, $T_{B}$, of individual
cotton strings using the setup shown in Fig. \ref{fig3:experiment}(a).
A bucket was suspended from a single string, and water was poured
slowly into the bucket until the string broke. The total weight of
the water and bucket at the breaking point was measured to obtain
$T_{B}$. Multiple measurements were performed to account for statistical
variations and potential defects in the strings. As shown in Fig.
\ref{fig3:experiment}(d), the measured breaking tensions exhibited
some scatter, with a mean value of $T_{B}=8.2\pm0.4$ N, represented
by the gray band (the value $\pm0.4$ N is calculated from the standard
deviation). We found that careful, slow pouring was essential to obtain
consistent results, and measurements yielding significantly lower
$T_{B}$ values (attributed to string defects) were excluded from
the analysis.

Next, we investigated the three-rope Y-configuration without mass
at the junction ($m=0$), as depicted in Fig. \ref{fig3:experiment}(b).
For various fixed angles $\theta$, water was gradually added to the
bucket until one of the strings broke. The angle $\theta$ was measured
precisely at the moment of string breaking. Our experiments confirmed
the theoretical prediction: for $\theta<\theta_{C}=60^{\circ}$, the
lower rope consistently broke first, while for $\theta>\theta_{C}$,
one of the upper ropes failed (with the specific upper rope breaking
depending on minor defects and little angular asymmetries). The measured
maximum exerted force, $T_{1}^{(B)}$, for $\theta>\theta_{C}$ is
plotted in Fig. \ref{fig3:experiment}(e), showing good agreement
with the theoretical expectation from Eq. \ref{eq:T1B}. The theoretical
expectation is obtained by using the value $T_{B}=8.2$ N in Eq. \ref{eq:T1B}.
The upper limit of the theoretical expectation is obtained by using
the upper limit of $T_{B}$, which is $8.2+0.4$ N, and the angle
$\theta\rightarrow\theta-\Delta\theta$ in Eq. \ref{eq:T1B} (note
that $\cos\theta$ is a decreasing function in this range). The lower
limit is obtained by using $T_{B}\rightarrow8.2-0.4$ N and $\theta\rightarrow\theta+\Delta\theta$
in Eq. \ref{eq:T1B}. The error of the angle measurement is considered
to be $\Delta\theta=1^{\circ}$. The area between the upper and lower
theoretical expectations is shaded gray. We note that due to the slight
elasticity of the cotton strings, the angle $\theta$ decreased gradually
during loading, requiring careful measurement at the instant at which
the strings break.

Finally, we examined the case with mass $m=0.400$ kg attached at
the junction point, using the same three-rope setup shown in Fig.
\ref{fig3:experiment}(c). According to Eq. \ref{eq:T1Bm}, this additional
mass reduces the critical angle to $\theta_{C,m}\simeq42^{\circ}$,
significantly lower than the mass-less case. Therefore, we focused
our measurements on angles below $60^{\circ}$ to clearly demonstrate
this shift. The experimental results for $T_{1}^{(B)}$, shown in
Fig. \ref{fig3:experiment}(f), agree well with the theoretical prediction
from Eq. \ref{eq:T1Bm} (gray band, obtained through a similar procedure
to the one of panel (e), but using Eq. \ref{eq:T1Bm}), confirming
that the presence of mass at the junction alters the breaking behavior
as predicted.

\section*{Summary}

In this work, we show that by studying the distribution of tension
and the conditions for rope failure in idealized multi-rope systems,
a deeper insight into Newton's law can be attained.

Beginning with a simple two-rope vertical configuration, where a body
of mass $m$ is sustained by a vertical tension $\overrightarrow{T}_{2}$
and is pulled downward by applying a tension $\overrightarrow{T}_{1}$
to it. In this case we show that the rate at which the pulling force
is applied, suddenly or gradually, determines whether the upper or
lower rope breaks. The same analysis is extended to a three-rope Y-configuration
system, for which we may identify a critical angle $\theta_{C}$ beyond
which the failure shifts from the lower to the upper ropes. We further
examine how the introduction of mass at the junction point alters
the critical angle and imposes mass-dependent limits on the system
stability.

These scenarios, though analyzed under ideal assumptions (mass-less,
rigid ropes and uniform breaking tension), mirror real-world behaviors
when adjusted for elasticity and damping. More importantly, they offer
opportunities to implement laboratory activities. In fact, the rope-breaking
phenomena can be reproduced and tested through hands-on experiments.
On their turn, these activities may foster inquiry, model-based reasoning,
and deeper understanding of Newtonian mechanics. To clearly demonstrate
one example of such activities, we conceived an experimental setup
and performed some measurements and showed their agreement with the
theoretical formulas within experimental errors.

Therefore, by blending analytical modeling with engaging experimental
activities, this work may serve as a resource for educators aiming
to connect elementary physics principles with active learning.

\bibliographystyle{apsrev4-2}
\bibliography{biblio}

\clearpage
\onecolumngrid
\noindent
\begin{minipage}{\textwidth}
\begin{minipage}{0.2\textwidth}
\centering
\includegraphics[width=0.9\linewidth]{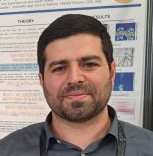}
\end{minipage}%
\hfill
\begin{minipage}{0.75\textwidth}
\textbf{Amir Eskandari-asl} \\
Amir Eskandari-asl, born on January 30, 1987, earned his Ph.D. in Condensed Matter Physics from Shahid Beheshti University, Tehran, Iran, in 2017. He is currently a postdoctoral researcher at the University of Salerno, Italy, where he works on theoretical condensed matter physics with a focus on ultrafast phenomena, light-matter interaction, and out-of-equilibrium quantum dynamics. He has extensive teaching experience at both undergraduate and graduate levels, including courses in Non-Equilibrium Physics, Solid State Physics, Mathematical Physics, and Basic Physics. He has authored several research papers in condensed matter theory and contributed to physics education through publications such as his work on the static properties of Slinky. His research contributions include the development of theoretical frameworks for understanding pumped quantum systems and quantum transport in nanoscale devices.
\end{minipage}

\vspace{0.5cm}

\begin{minipage}{0.2\textwidth}
\centering
\includegraphics[width=0.9\linewidth]{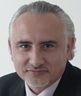}
\end{minipage}%
\hfill
\begin{minipage}{0.75\textwidth}
\textbf{Roberto De Luca} \\
Roberto De Luca, born on June 8, 1958, earned a degree in physics from the University of Salerno, Italy, in 1986. He received his MA in physics from the University of Southern California in 1987 and his PhD from the University of Naples and Salerno in 1992. Since 1994, he has been conducting research at the University of Salerno.
His original research focused on Josephson junction array models and the dynamics of superconducting devices. Since 2001, he has authored papers on physics education. Some of his works in this field have been dedicated to finding simple ways to introduce alternative energy concepts to young students.
\end{minipage}
\end{minipage}

\twocolumngrid


\end{document}